***************************************************************************
\documentstyle[twocolumn,aps]{revtex}
\def\r{{\bf r}}
\def\x{{\bf x}}

\def\n{{\bf n}}

\def\u{{\bf u}}
\def\ttheta{{\tilde\theta}}
\def\nab{{\bf \nabla}}
\def\begineq{\begin{equation}}
\def\endeq{\end{equation}}

\newcounter{saveeqn}
\newcommand{\alpheqn}%
{\setcounter{saveeqn}{\value{equation}}\setcounter{equation}{0}%
\renewcommand{\theequation}%
{\mbox{\arabic{saveeqn}\alph{equation}}}%
\addtocounter{saveeqn}{1}}
\newcommand{\reseteqn}{\setcounter{equation}{\value{saveeqn}}%
\renewcommand{\theequation}{\arabic{equation}}}
\renewcommand{\theequation}{\arabic{equation}}

\begin{document}
\preprint{Preprint}

\title{Fractal dimension crossovers in turbulent passive scalar signals}
\author{Siegfried Grossmann and Detlef Lohse}

\address{ Fachbereich Physik, Philipps Universit\"at,
 Renthof 6, D-35032 Marburg, FRG}
\date{\today}

\maketitle

\begin{abstract}
The fractal dimension $\delta_g^{(1)}$ of turbulent passive  scalar
signals  is  calculated from the fluid dynamical equation.
$\delta_g^{(1)}$ depends on the scale. For small Prandtl
(or Schmidt)
number  $Pr<10^{-2}$  one gets two ranges,  $\delta_g^{(1)}=1$  for  small
scale r and $\delta_g^{(1)}$=5/3 for large r, both as expected.
But for large $Pr> 1$
one gets a third,  intermediate range in which the signal is extremely wrinkled
and has $\delta_g^{(1)}=2$. In that range
the  passive  scalar
structure function $D_\theta(r)$ has a plateau.
We calculate the $Pr$-dependence  of   the
crossovers. Comparison with a numerical reduced wave vector set calculation
gives good agreement with our predictions.
\end{abstract}

\pacs{45-11c}

%\begin{narrowtext}
\narrowtext
%-----------------------------------------------------------------------
As seen in recent experiments \cite{til93,pro91}
the temperature power spectrum
for thermally  driven turbulence shows a strong dependence
on the  Prandtl
number
$Pr=\nu/\kappa$. $\nu$ is the viscosity of the fluid and  $\kappa$
the thermal conductivity (diffusivity) of the advected scalar.
While for the  helium  cell  ($Pr\approx   0.7$)
there  was a large scaling  range
\cite{pro91}, no  universal
scaling could be found for water ($Pr\approx 7$) \cite{til93}.

We took these experiments as a motivation to examine  the $Pr$
dependence of
self similarity    features   of   a   passive   scalar    field
$\theta(\x,t)$.
The   passive  scalar  $\theta$  could  be  the
temperature or a dye (then $Pr$ is often denoted as Schmidt number $Sc$),
convected by a turbulent, isotropic
velocity field $\u (\x,t)$,
\begineq
\partial_t \theta = - \u \cdot \nab \theta +\kappa {\nab}^2
\theta + f_\theta.
\label{eq1}
\endeq
$f_\theta(\x,t)$  is  a  forcing  term  replacing the  boundary
conditions.

Self similarity in turbulence is commonly characterized
by the scaling exponents of power
spectra  or  of structure functions. Alternatively, one may consider
the  fractal
dimensions  $\delta_g^{(d)}$  of the  d-dimensional  graphs  of
hydrodynamic fields. For the passive scalar $\theta (\x ,t)$ the
scaling  exponents  $\zeta_m^{(\theta)}$  of  the  structure
functions are defined by
\begineq
\langle |\theta (\x +\r ,t) - \theta (\x ,t ) |^m \rangle
\propto r^{\zeta_m^{(\theta)}}.
\label{eq2}
\endeq

The scale dependence of
the  Hausdorff volume  $H^{(d)}(G(B_r^{(d)}))$
of the graph
$G(B_r^{(d)}) = \left\{ (\x,\theta) |\x \in B_r^{(d)},
\theta=\theta(\x ) \right\}$
over a ball $B_r^{(d)}$ of radius r defines the fractal dimension
 $\delta_g^{(d)}$,
\begineq
H^{(d)}(G(B^{(d)}_r)) \propto
 r^{\delta_g^{(d)}}.
\label{eq3}
\endeq

In particular, $\delta_g^{(3)}$ is the Hausdorff dimension of the
passive  scalar  graph over a  three  dimensional  ball
$B_r^{(3)}$, and $\delta_g^{(1)} =\delta_g^{(3)}-2$ \cite{con93}
is the Hausdorff dimension of a turbulent signal in  space
(for   fixed   time)  or  --  by  the   Taylor   hypothesis
\cite{Tay38} -- in time for fixed position.

For the passive   scalar   the    scaling    exponent
$\zeta_1^{(\theta)}$ of the structure function
and the fractal dimension $\delta_g^{(d)}$ are
connected by
\begineq
\delta_g^{(d)} \le d +(1-\zeta_1^{(\theta)}).
\label{eq4}
\endeq
We suppose as in \cite{con93}
that the inequality is in fact sharp.

In  this  paper we shall calculate the  fractal  dimension
$\delta_g^{(3)}$  and  thus via (\ref{eq4}) also the  scaling  exponent
$\zeta_1^{(\theta)}$ {\it from the dynamical equation} (\ref{eq1}).
The main tool of our calculation is the  volume  formula
for  the passive scalar graphs \cite{Fed69},  which  was
introduced  as a very useful tool into fluid dynamics  by
Constantin and Procaccia \cite{Con90,con93,Pro93a}.
Its main  advantage is that  one  can  apply
rigorous techniques with controlled approximations,  based
on  the  fluid  dynamical  equations.   By  extending  the
calculations of \cite{con93,fai93} we are able to go
beyond an estimate of the exponents: we handle also the amplitudes.
Doing so, we confirm the various scaling ranges addressed in \cite{fai93}
and calculate the  $Pr$
dependence of the crossovers. Comparison
with experiments, simulations, and former theories is discussed.

For  convenience,  we measure the passive scalar  field  in
multiples of its rms,  $\tilde\theta=\theta/\theta_{rms}$.
We  do  not  need to assume that upper  bounds
$\theta_{max}$  or  $u_{max}$  for the  passive  scalar  or
velocity field exist, loosening thus the assumptions made in \cite{con93}.
We only need ${\cal L}_2$-integrability which anyhow is necessary for the
existance of structure functions.

According to geometric  measure theory \cite{Fed69,con93}  the
Hausdorff   volume   $H^{(3)}(G(B_r^{(3)}))$   of   the   graph
$r\tilde\theta(\x)$ over the ball $B_r^{(3)}=:B_r$ reads
\begineq
H^{(3)}(G(B_r)) = \int_{B_r} d^3x \sqrt{1+ r^2 |{\nab}^2 \tilde\theta |^2}.
\label{eq5}
\endeq
In one dimension eq.(\ref{eq5}) is the well known formula for  the
length of the  curve $r\tilde\theta(x)$.
Dividing by $V^{(3)}(B_r)=4\pi r^3/3$ and applying
Cauchy-Schwarz's inequality we get
\begineq
H^{(3)}(G(B_r))/
V^{(3)}(B_r)
\le \sqrt{  1+ {3\over 4\pi r} \int_{B_r} d^3x
|{\nab}^2 \tilde\theta |^2}.
\label{eq6}
\endeq

Now  the nice idea of \cite{con93} was to calculate
$|\nab \theta|^{2}$ from
the heat transfer equation (\ref{eq1}). In the stationary case it is
\begineq
|\nab\ttheta|^2={1\over 2\kappa } \left\{-\u\cdot\nab + \kappa\nab^2
\right\} \ttheta^2 + {f_\theta\ttheta\over \kappa \theta_{rms}}.
\label{eq7}
\endeq
We insert the three terms of (\ref{eq7}) in (\ref{eq6}) and denote the
resulting terms under the square root in (\ref{eq6}) by
$I_1$, $I_2$, and $I_3$, respectively. Directly from the
definition of  the thermal intensity dissipation rate $\epsilon_\theta$, we
get $I_3 = {\epsilon_\theta\eta^2\over \kappa\theta_{rms}^2}\left(
{r\over\eta}\right)^2$. Here $\eta=\nu^{3/4}/\epsilon^{1/4}$ is the Kolmogorov
length and $\epsilon$ the energy dissipation rate \cite{my75}. $I_3$ can
further
be estimated by $I_3 \sim Pr Re^{-1/2} (r/\eta )^2$. Similarily, $I_2$ can be
bound by $I_2 \le \sqrt{3I_3} \sim Pr^{1/2} Re^{-1/4} r/\eta$.

For sufficiently large $Re$, $I_2$ and $I_3$ can be neglected for all relevant
r, i.e., for r smaller than the outer length scale $L$.
We are  left
with  $I_1$. Applying Gauss' theorem, we get
\alpheqn
\begineq
I_1 =
{3\over 8\pi\kappa r} \oint_{\partial B_r} \ttheta^2 (\x)\u (\x)\cdot \n(\x)
dA(\x).
\label{eq8a}
\endeq
$\n(\x)$  is  the unit vector normal to  the  sphere,  directed
inwards.
The  central idea of \cite{con93} now  was  to
introduce  velocity  and scalar field {\it differences} in (\ref{eq8a})
to  connect  the
Hausdorff  dimension  of the passive scalar  with  the
r-scaling exponents of velocity and scalar differences.

Here this is achieved in a slightly different way by adding a term
$\propto \u ({\bf x_0} )$
%\cdot \oint_{\partial B_r} \tilde \theta^2 (\x )
%\n (\x ) dA (\x )$
to  the rhs of (\ref{eq8a}),  where $\x_0$ is the center of  $B_r$.
On average $\langle \u (\x_0)\rangle =0$.
Thus we are
allowed to write
\begineq
I_1=
{3r\over 2\kappa } \oint_{\partial B_r} \ttheta^2 (\x)
(\u (\x)-\u(\x_0)) \cdot \n(\x)
{dA(\x)\over A_r},
\label{eq8b}
\endeq
where  $A_r=4\pi r^{2}$ is the surface of  the  sphere.  We
again apply Cauchy-Schwarz and get
\begineq
I_1  \le
{3r\over 2\kappa } \sqrt{ \left\langle \ttheta^4 (\x)
\right\rangle_{\partial B_r}
\left\langle
\left((\u (\x)-\u(\x_0)) \cdot \n(\x)\right)^2
\right\rangle_{\partial B_r}}.
\label{eq8c}
\endeq
\reseteqn
$\langle \dots\rangle_{\partial B_r}$ denotes the averaging over the sphere.
The first factor under the square root is the flatness of the scalar
field, which is known to be 3 from experiment \cite{my75}. The second factor
is the {\it longitudinal}
velocity structure function $D_{\|}(r)$.

$D_{\|}(r)$  is related to the  velocity  structure
function
$D(r)=\langle |\u(\x+\r) -\u(\x)|^2\rangle$ via incompressibility
\cite{my75},
\begineq
D_{\|}(r) =r^{-3} \int_0^r \rho^2 D(\rho) d\rho.
\label{eq9}
\endeq
If  we  measure  lengths in multiples  of
$\eta=\nu^{3/4}/\epsilon^{1/4}$,
and   velocities  in  multiples   of   the
Kolmogorov  velocity  $v_\eta=(\nu\epsilon)^{1/4}$
\cite{my75},  $\tilde D=D/v_\eta^2$,
$\tilde r=r/\eta$, we finally obtain
\begineq
{H^{(3)}(G(B_r))\over V^{(3)}(B_r)} =
\hbox{const}~
r^{\delta_g^{(3)}-3}
\le \sqrt{1+ {3\sqrt{3}\over 2} Pr \tilde r \sqrt{\tilde D_{\|}(\tilde r)}},
\label{eq10}
\endeq
which is our main result.  The inequalities arise from the
Cauchy-Schwarz  estimation. It is thus  reasonable  to
assume  that at least the $r$- and $Pr$-scaling  behaviour  is
correctly  given  by  (\ref{eq10}).  We  thus  have  a  controlled
bound for  the volume $H^{(3)}(G(B_r))$  of  the
passive scalar graphs, {\it including} the amplitudes.

Now  from  experiment  it  is  known  that  the  Batchelor
interpolation formula is an excellent fit for the velocity
structure function \cite{my75},
\begineq
\tilde D (\tilde r) ={\tilde r^2 \over 3 (1+a^2 \tilde r^2 )^{2/3}},
\qquad a^{-1} =11.2.
\label{eq11}
\endeq

We  use  the  interpolation  formula (\ref{eq11}) to  calculate  the
fractal  dimension $\delta_g^{(3)}$ from (\ref{eq10}) and  (\ref{eq9})
and get
\begineq
\delta_g^{(3)}-3 =
{d\over d\ln\tilde r} \ln
 \sqrt{1+ {3\sqrt{3}\over 2} Pr \tilde r
 \sqrt{\tilde D_{\|}(\tilde r)} }.
\label{eq12}
\endeq

The  numerical  result from (\ref{eq12})
for $\delta_g^{(1)} = \delta_g^{(3)}-2 $
is given for  several  $Pr$
numbers  in Fig.1.  In principle,  four cases are  possible
which we now want to discuss.

For ``small'' $Pr$ there are two ranges.  If r is sufficiently small,
we  have
$\delta_g^{(1)}=1$  and  from (\ref{eq4})  $\zeta_1^{(\theta)}=1$.  On  these
small  scales both the scalar and the velocity  field  are
smooth,   $\tilde D_{\|}(\tilde r)$  is  given  by
$\tilde D_{\|}(\tilde r)=\tilde r^{2}/15$.  For
increasing  r  a  crossover occurs in  $D_{\|}(r)$  and  for
$r>14\eta$ the longitudinal structure function is given by
$\tilde D_{\|}(\tilde r)=(3/11)b\tilde r^{\zeta_2^{(u)}}$
with \cite{my75} $b=a^{-4/3}/3=8.4$,
$\zeta_2^{(u)}$ near $2/3$.
The  velocity  field is now fractal but, as $Pr$ is considerd to be small,
no change  can  be
observed  in  the fractal dimension $\delta_g^{(1)}=1$  of  the
passive  scalar field which stays to be smooth,  as
the 1 under the square root in (10) is still dominant.
Physically this means that the diffusivity $\kappa$ is  so
large  that  the passive scalar field is smooth  even  on
turbulent scales. For sufficiently large scale,
\begineq
r/\eta \ge \left( {2\over 3\sqrt{3}} \sqrt{{11\over 3b}}\right)^{3/4}
 Pr^{-3/4} = 0.36 Pr^{-3/4}
\label{eq13}
\endeq
the   second   term  in  (\ref{eq10})   becomes   dominant.   Then
$\delta_g^{(1)}=3/2+\zeta_2^{(u)}/4$   and  from   (\ref{eq4})
$\zeta_1^{(\theta)}=1/2-\zeta_2^{(u)}/4$.     Without    the    at most tiny
intermittency  corrections  we  have  $\delta_g^{(1)}=5/3$  and
$\zeta_1^{(\theta)}=1/3$.  Here  both  velocity  and  passive
scalar  field are fractal and scale alike.  The  classical
Obukhov-Corrsin scaling theory \cite{obu49} is recovered.

To  observe  the transition (\ref{eq13}) we  must  have
$0.36Pr^{-3/4}>14$,  which implies the condition
$Pr<10^{-2}$. Thus ``small'' $Pr$ are those with $Pr < Pr_l\approx 10^{-2}$.
Furthermore, if $Pr$ is even smaller than $(4Re)^{-1}$, the classical
Obukhov-Corrsin scaling range can never be achieved, because $r<L$, i.e., the
temperature signal is smooth on all scales.

For  ``large''  $Pr$  the situation is alike for sufficiently small  and
sufficiently large r.  But for intermediate r the second term under the
square   root  in  (\ref{eq10})  can  become   already dominant   although
still $r<14\eta$, i.e., still   in  the viscous
range of the  velocity  field,
if only $Pr$   is   large enough.    In this range $\delta_g^{(1)}=2$    and
consequently $\zeta_1^{(\theta)}=0$.  This means that
the  passive scalar signal is  highly
wrinkled although the velocity field is completely smooth
on that scales.  In that situation the dye or the heat  is
very  efficiently  mixed  by the
velocity  field which is advected by the larger turbulent   eddies.
But  since  the diffusion is very  slow
(large $Pr$ means small  $\kappa$), concentration or temperature differences
cannot be smeared out. Similar phenomena are obtained when
non-turbulent,  viscous fluids are mixed:  fractal patterns
develop \cite{ott88}.

{}From (\ref{eq10}) we calculate that this intermediate range begins
at
\begineq
r/\eta \le \sqrt{2 \sqrt{5}/3}
 Pr^{-1/4} = 1.22 Pr^{-1/2}.
\label{eq14}
\endeq

It ends at $r/\eta=14$. Thus, to develop such an
intermediate range of say, a decade, we must have
$Pr >  Pr_u \approx 1$.
Such $Pr$ we denote as "large".

Let  us  now have a look on the passive  scalar  structure
function $D_\theta(r)= \langle |\theta (\x +\r ) - \theta (\x )|^2\rangle
\propto r^{\zeta_2^{(\theta )}}$
Neglecting   again possible  intermittency    corrections,    we    have
$\zeta_2^{(\theta)}=2\zeta_1^{(\theta )}=4-2\delta_g^{(1)}$.      For
small  r  and large r we  have  $\zeta_2^{(\theta)}=2$  and
$\zeta_2^{(\theta)}=2/3$,  respectively,  as for the velocity
structure   function.   But,   as   derived   above,   for
intermediate  r  in the case of large $Pr$ we have a  plateau  in  the
passive scalar structure  function,  $\zeta_2^{(\theta )}=0$,
$D_\theta(r)=$ const.
The complete structure function can easily
be reconstructed from the scale dependent
scaling exponent $\zeta_2^{(\theta )}(r)=2-\delta_g^{(1)}(r)$.
The result is shown is Fig.2.

The  plateau has already been predicted by a  mean  field
theory  \cite{Eff89}.  In that theory it is observed  also
for $Pr>Pr_u$ with a very similar $Pr_u$,  see Fig.5
of \cite{Eff89}. The extension
of  the intermediate range here,  $1.22Pr^{-1/2}<r/\eta<14$,  is
somewhat   different   from
$5.49Pr^{-1/4}<r/\eta<15.6Pr^{3/4}$ in \cite{Eff89}, probably due to
the mean field approximations in \cite{Eff89}.

The theory of Batchelor \cite{bat59,my75} also predicts  an
intermediate range for $Pr >1$.  In that theory $D_\theta (r)$
depends  only  logarithmically on r  in  the  intermediate
range   $r_1<r<r_2$, but with   $r_1\propto Pr^{-1/2}$   and   $r_2$
independent of $Pr$ as in our theory.

For small $Pr <  Pr_l$ all three theories predict the  transition
{}from  $\zeta_1^{(\theta)}=1$ to $\zeta_1^{(\theta)}=1/3$  for  a
scale $\propto  Pr^{-3/4} \eta$,  see eq.(\ref{eq13}),  refs.\
\cite{Eff89},  and
\cite{my75}.

Note that a plateau in the temperature structure function means a
$k^{-1}$-behavior in the temperature spectrum. This $k^{-1}$-behavior was
also postulated by Kraichnan \cite{kra74}.

Experiments and full numerical simulations for
large Reynolds numbers $Re$ {\it and} large Prandtl numbers $Pr$ are very rare.
There are some hints that there in fact {is} a plateau, see the collection
of experimental data in \cite{my75}.
To get independent confirmation of our predictions, we numerically solved
eq.\ (\ref{eq1}) together with the Navier-Stokes equation in a reduced wave
vector set
calculation for several $Pr$. For details concerning the method, see
refs.\ \cite{gnlo91,gnlo93e}. The result is shown in Fig.3.
For small wavevector $p$, i.e., large $r$, there is classical scaling,
$\zeta_2^{(\theta)}= \zeta_2^{(u)}=2/3$, for all scales. But
for large $Pr>1$ a plateau develops for
medium scales as predicted by our theory.
For small scales (large $p$) the spectra fall off exponentially, which
reflects that the signal is smooth ($\delta_g^{(1)}=1$) on these scales.

It becomes particularly evident from our figures, that
$\delta_g^{(1)}$ and $\zeta_2^{(\theta)}=4-
2\delta_g^{(1)}$ should not be considered as {\it global} scaling
exponent,    but, instead, as   {\it local}   scaling    exponents
$\delta_g^{(1)}(r)$ and
$\zeta_2^{(\theta)}(r)$, a concept which we have also introduced
to examine the intermittency corrections in the Navier-Stokes dynamics
\cite{gnlo93e}.
If    the    r-ranges,    where
$\zeta_2^{(\theta)}(r)$ stays at a certain value, are small,  it will be
difficult to extract scaling exponents from experimental
or simulated data.

Here this situation appears for $Pr>Pr_u\approx 1$.
This might explain one aspect of the
above  mentioned Rayleigh-Benard experiments of  Libchaber
and  coworkers  where no global scaling exponent can be found
for  the ``large'' $Pr$ number $ Pr\approx 7$
\cite{til93}.  Note that our theory can  also  be
applied   for  an  active  scalar  as  for   example   the
temperature field in Rayleigh-Benard convection. Eq.(\ref{eq10}) remains
valid for an active scalar,  but, instead of the Batchelor
interpolation  (\ref{eq11}),  the
velocity structure function of a {\it thermally} driven velocity
field has to be used,  which unfortunately is still not  known reliably.
If we assume Bolgiano-Obukhov scaling \cite{Bol59} for the
velocity  field,  $D(r)\propto  r^{6/5}$,  we also  get
Bolgiano-Obukhov     scaling    for    the    temperature     field,
$D_\theta(r) \propto r^{2/5}$, if only $r$ is sufficiently large.

The existance of {\it non-global}
scaling ranges is also predicted for
low Prandtl number Rayleigh-Benard convection, but
only for very high Rayleigh numbers \cite{gro93}.

After having introduced {\it global} scaling as a paradigma in
nonlinear dynamics, we might have to get used to {\it non-global}
scaling ranges as the {\it normal} case, where
universal exponents can hardly be derived from experiments or simulations,
as the individual scaling ranges are too small.

%\end{narrowtext}

\acknowledgements
Partial support by GIF is gratefully acknowledged.
The   HLRZ  J\"ulich
supplied us with computer time.

\vspace{1cm}

\begin{figure}
\caption{Scale dependent fractal dimension $\delta_g^{(1)}$ of the
passive scalar signal for several $Pr$.}
\end{figure}

\begin{figure}
\caption{Temperature structure function
$\tilde D_\theta (\tilde r)=
D_\theta (\tilde r)/(\epsilon_\theta\epsilon^{-1/2}\nu^{1/2})$.
The plateau can easily be recognized for sufficiently large $Pr$.
$\log \tilde D_\theta (\tilde r)$ is obtained by numerical integration
of $ \zeta_2^{(u)} (\tilde r)$
over $\log\tilde r$.}
\end{figure}

\begin{figure}
\caption{Temperatur spectra from an approximate solution of the
dynamical equations (by a  reduced wave vector
set method)
%\cite{gnlo91,gnlo92a,gnlo92b,gnlo93e}
for several $Pr$. The Reynolds number always is $Re=2\cdot 10^4$.
The dashed-dotted line denotes the velocity spectrum of the advecting fluid.
$k^{-\zeta}$-behavior in the calculation here with discrete wave vectors
means
$k^{-\zeta -1}$-behavior in a continuous wave vector calculation.
}
\end{figure}

\end{document}